\documentstyle[12pt]{article}

\begin{document}
\tolerance=5000
\def\be{\begin{equation}}
\def\ee{\end{equation}}
\def\bea{\begin{eqnarray}}
\def\eea{\end{eqnarray}}
\def\nn{\nonumber \\}
\def\e{{\rm e}}

\makeatletter
\renewcommand{\theequation}{\thesection.\arabic{equation}}
\@addtoreset{equation}{section}
\makeatother
\renewcommand{\thefootnote}{\fnsymbol{footnote}}

\thispagestyle{empty}

\vspace*{-20mm}

\  \hfill
\begin{minipage}{3.5cm}
YITP-04-37 \\
July 2004 \\
hep-th/0407083 \\
\end{minipage}

\vskip 2cm

\begin{center}

{\Large\bf Holography
\vspace{3mm}

 in Einstein Gauss-Bonnet Gravity}
\end{center}

\vskip 2 cm
\begin{center}
{\bf Sachiko Ogushi\footnote{E-mail: ogushi@yukawa.kyoto-u.ac.jp}
and Misao Sasaki\footnote{E-mail: misao@yukawa.kyoto-u.ac.jp}}
\\
\vskip 0.2 cm
{\it Yukawa Institute for Theoretical Physics,}
\\
{\it Kyoto University, Kyoto 606-8502, Japan}

\vskip 15mm

{\bf ABSTRACT}
\end{center}

We investigate a holographic relation between
Einstein Gauss-Bonnet gravity in $n$ dimensions and its dual
field theory in ($n-1$) dimensions.
We briefly review the AdS/CFT correspondence for the entropy in
the $n$-dimensional Einstein gravity and consider its extension
to the case of the $n$-dimensional Einstein Gauss-Bonnet gravity.
We show that there is a holographic relation between entropies of 
an Einstein Gauss-Bonnet black hole in the bulk and the corresponding
radiation on the brane in the high temperature limit.
In particular, we find that the Hubble entropy evaluated
when the brane crosses the horizon also coincides with
the black hole entropy in the high temperature limit.

\newpage

\setcounter{page}{1}

\setcounter{footnote}{0}
\renewcommand{\thefootnote}{\arabic{footnote}}

\section{Introduction}

The idea that our universe is a (mem)brane in a higher dimensional
spacetime and all the interactions but gravity are confined on the brane,
the so-called braneworld scenario, has attracted much attention
in recent years. Among others, one of the most intriguing models of 
the braneworld was proposed by Randall and Sundrum~\cite{RS2},
because Einstein gravity is recovered on the brane in the large
distance limit. In this model, a single positive tension brane is 
embedded in the 5-dimensional anti-de Sitter space (AdS$_5$) with
$Z_2$ symmetry. It was then soon realized that a cosmological version
of this model can be regarded as a physical realization of the
AdS/CFT correspondence~\cite{Gubser:1999vj,GS,SV,GP1,GP2,GP3},
which otherwise describes a formal, mathematical correspondence between
gravity in the AdS bulk and a CFT on the brane at the boundary of 
AdS~\cite{Maldacena:1997re}.

Specifically, the AdS/CFT correspondence relates thermodynamics of 
an ${\cal N}=4$ super Yang-Mills theory in 4-dimensions to
thermodynamics of Schwarzschild black holes in AdS$_5$.
It is argued by Witten~\cite{EW} that the entropy of a CFT at
high temperature can be identified with the entropy 
of an AdS black hole. As we review in the next section, the high
temperature limit means the case when the horizon radius is much
greater than the AdS curvature radius. Thus, in this limit,
the fact that the black hole entropy is proportional to the horizon
area is understood as a result of holography,
i.e., the degrees of freedom of the AdS Schwarzschild bulk are
proportional to the area of its boundary.

An interesting, non-trivial relation between a CFT and 
a radiation-dominated Friedmann-Lematre-Robertson-Walker (FLRW) universe
was founded by Verlinde~\cite{EV}. There he generalized 
the entropy formula of the well-known Cardy formula~\cite{CA} for
a CFT in two-dimensions to arbitrary dimensions.
The so-called Hubble entropy plays a central role in connecting
the CFT entropy formula with the FLRW equation.
The ($n-1$)-dimensional Hubble entropy is defined by
\bea
\label{sn5}
S^{\rm H}_{n-1}= {(n-3) H V(r) \over 4 G_{n-1}} \;,
\quad V(r)= r^{n-2} \Omega_{n-2} \; .
\eea
where $V$ is the volume of ($n-2$)-dimensional sphere of radius $r$.
Savonije and Verlinde considered a brane-universe embedded in an 
$n$-dimensional AdS Schwarzschild space and showed the
correspondence between the entropy of CFT and the FLRW equation
explicitly~\cite{SV}. Namely, they showed that the Hubble entropy 
and the black hole entropy coincides when the brane crosses the 
black hole horizon,
and the CFT entropy formula reproduces the FLRW equation
if the volume $V$ in Eq.~(\ref{sn5}) is fixed at $V=V(r_{+})$,
where $r_{+}$ is the horizon radius. 

All the results mentioned above are based on the assumption, either
explicitly or implicitly, that gravity in the bulk is described by Einstein.
It is therefore natural to ask if the same or similar results
hold for a more general theory of gravity.
In this paper, as one of such generalizations, we consider
the case of the Einstein Gauss-Bonnet theory~\cite{DL}.
The Einstein Gauss-Bonnet theory has a special combination
of curvature-squared terms, the Gauss-Bonnet term,
added to the Einstein-Hilbert action.
The Gauss-Bonnet term is given by
\bea
\label{gbt}
{\cal L}_{\rm GB}=R_{abcd}R^{abcd}-4R_{ab}R^{ab}+R^2 \; .
\eea   
Note that in $n=4$ dimensions, 
the Gauss-Bonnet term is a topological invariant 
that does not enter the dynamics. We therefore consider $n \ge 5$.
The Einstein Gauss-Bonnet theory is a simplest natural extension
of the Einstein theory in the sense that no derivatives higher than
second will appear in the field equation.
In the context of string theory, this term
plays a role of the leading order stringy corrections to Einstein gravity
in the heterotic string~\cite{CH}.
Moreover the Gauss-Bonnet combination of curvature-squared terms
is needed to cancel the ghost term in the low energy supergravity
limit of the heterotic string~\cite{BZ,BZU,GSl}.  
{}From the viewpoint of AdS/CFT correspondence, it is argued that
the Gauss-Bonnet term in the bulk corresponds to next to leading order
corrections in $1/N$ expansion of a CFT~\cite{NO}. 

Turning our eyes on thermodynamic quantities associated with
an Einstein Gauss-Bonnet black hole, there exist notable differences
in comparison with the case of Einstein gravity. 
In particular, the entropy of an Einstein Gauss-Bonnet black hole 
does not obey the area law~\cite{MS}, though the violation is
small at high temperatures. Detailed discussions on thermodynamics of 
Einstein Gauss-Bonnet black holes can be found
in~\cite{SN1,CAI,SN2,NP2,CAI2,CAI3}.
Thus it is interesting to investigate if a holographic relation can
 still persist in some
way or other between the Einstein Gauss-Bonnet bulk and its dual field
theory that includes $1/N$ corrections to CFT at finite temperatures.
There have been a number of works related to this
 topic~\cite{SN3,SN4,NP,SN5,CAI1,GP}.
In Ref.~\cite{GP}, Gregory and Padilla pointed out that there is a 
holographic relation for a critical brane in the Gauss-Bonnet gravity 
in the limit of the brane at infinite boundary of AdS at high temperatures,
but there is no correspondence of the Cardy-Verlinde formula 
through Eq.~(\ref{sn5}).

In this paper, following the method used in \cite{GS} and \cite{SV}, 
we first describe our method by showing the AdS/CFT correspondence
for the entropy in $n$-dimensional Einstein gravity,
and apply it to $n$-dimensional Einstein Gauss-Bonnet gravity.
We show that there is a holographic relation in 
Einstein Gauss-Bonnet gravity for the brane near the AdS boundary
and at high temperatures. Then, rather unexpectedly, we find in the
high temperature limit that the Hubble entropy~(\ref{sn5}) evaluated
at the horizon crossing agrees with the black hole entropy.
This indicates that a part of $1/N$ corrections to CFT simply contributes
to renormalization of some quantities at leading order. 
However, the Cardy-Verlinde formula that relates the CFT entropy with
the FLRW equation fails to hold, indicating the existence of non-trivial
$1/N$ corrections at next-to-leading order.

\section{Entropy correspondence at high temperature}

\subsection{Einstein Gravity}

We consider an ($n-1$)-dimensional brane with a positive tension
in the background of the $n$-dimensional AdS Schwarzschild black hole.
We assume $Z_2$ symmetry with respect to the brane, so that the bulk
spacetime ${\cal M}$ consists of two identical copies of the AdS 
Schwarzschild geometries glued at an $(n-1)$-dimensional timelike
hypersurface $\Sigma$ where the brane is located.
The total action of the system is given by
\bea
\label{ac0}
I&=&I_{\mbox{\scriptsize bulk}}+I_{\mbox{\scriptsize brane}} \nn
\label{ac1}
I_{\mbox{\scriptsize bulk}}&=&{1\over 16\pi G_n} \int_{\cal M} d^{n} x\sqrt{-g}
~ (R- 2 \Lambda_n ) \; ,\\
\label{ac2}
I_{\mbox{\scriptsize brane}}&=& 
-\int_\Sigma d^{n-1} x\,\sqrt{-h} ~\sigma
\; .
\eea
Here $g$ and $h$ are the determinants of the bulk and induced metric,
respectively, $G_n$ is the $n$-dimensional Newton constant,
and $\sigma$ is the tension of the brane.

The $n$-dimensional AdS black hole metric is given by 
\bea
\label{ein}
ds_n^2 &=&-f(r) dt^2 +{dr^2 \over f(r)}+r^2 d\Omega^2_{n-2}\; , \\
f(r)   &=& 1+{r^2 \over l^2}-{\mu \over r^{n-3}}.
\eea
Here $d\Omega_{n-2}$ is the metric
on a unit $(n-2)$-sphere and $\mu$ is the mass parameter which 
is related to the black hole mass $M$ by
\bea
\label{mass}
M ={(n-2)\Omega_{n-2}\mu \over 16 \pi G_n}
\eea
where $\Omega_{n-2}$ is the volume of the unit ($n-2$)-sphere.
$l$ is the curvature radius of the $n$-dimensional AdS which is 
related to the cosmological constant:
\bea
\Lambda_n= -{1\over 2 l^2}(n-1)(n-2)
\eea

One way to obtain the black hole entropy is to use 
a Euclidean method~\cite{GibHaw77,HawPag83}.
The Euclidean method is summarized as follows:
First, we calculate the Euclidean action $I$ on-shell.
Second, we calculate the energy $E$ from the
action $I$ as $E={\partial I/\partial \beta}$ where
 $\beta=1/T_H$.\footnote{The black hole mass $M$ in Eq.~(\ref{mass}) 
is equal to this energy $E$. There are also several other methods
to calculate $E$ which gives $E=M$~\cite{ADM,AD,KBL}.}
Third, we obtain the entropy $S$ from $S=\beta E-I$. 
A detailed derivation can be found in Ref.~\cite{EW}.
The entropy of an AdS Schwarzschild black hole
takes the same form as that of a Schwarzschild black hole,
i.e., it is given by the area of the horizon $r_+$:
\bea
\label{sn}
S_{n}={\mbox{Area} \over 4G_n}={\Omega_{n-2} r_{+}^{n-2} \over 4G_n}
\eea
{}From the point of view of the AdS/CFT correspondence,
we are interested in the holographic relation of the entropies 
between the $n$-dimensional AdS black hole and the ($n-1$)-dimensional
dual field theory living on its boundary.
The relation of Newton constants between $G_{n}$ and $G_{n-1}$ is known 
as follows~\cite{GKR}:
\bea
\label{ree}
G_n= {2 l \over n-3} G_{n-1}
\eea
Then if we express the $n$-dimensional entropy 
in terms of ($n-1$)-dimensional words,
the entropy takes the following form.
\bea
\label{sn1}
S_{n}={\Omega_{n-2} r_{+}^{n-2} (n-3) \over 8 l G_{n-1} }\,.
\eea

Our aim is to compare the above $n$-dimensional entropy with 
the ($n-1$)-dimensional entropy on the brane.  For this purpose, 
we consider the dynamics of the brane in the bulk.
We parametrize the brane trajectory in the bulk 
by $(t,r)=(t(\tau),r(\tau))$.
The induced metric on the brane is
\bea
\label{frw}
ds_{n-1}&=&h_{ab}dx^{a}dx^{b} \nn
&=&-d\tau^2+r(\tau)^2 d \Omega_{n-2}^2 \;,
\eea
where the parameter $\tau$ is taken to be the proper
time on the brane. This implies the following condition:
\bea
\label{cond}
-f(r) \dot{t}^2 + {\dot{r}^2 \over f(r)}=-1\,,
\eea
where the overdot denotes the differentiation with respect to $\tau$.
The equation of motion for the brane is determined by the
junction condition as
\bea
\label{eqk}
2({\cal K}_{ab}-{\cal K}h_{ab} )= -8\pi G_n S_{ab}\,,
\eea
where the $Z_2$ symmetry is assumed, ${\cal K}_{ab}$ is 
the extrinsic curvature given by ${\cal K}_{ab}=h^{c}_{a}\nabla_{c}n_{b}$ 
where $n_{a}=(\dot{r},-\dot{t},0,\cdots,0)$ is the unit normal to the brane,
and $S_{ab}$ is the energy momentum tensor derived from
 $I_{\mbox{\scriptsize brane}}$.  In the present case of a vacuum brane,
we have $S_{ab}=-\sigma\, h_{ab}$. 
The angular component of Eq.~(\ref{eqk}) reads 
\bea
\label{eqk1}
f(r){\dot{t}\over r}={4\pi G_n \over (n-2)} \sigma\; .
\eea
Substituting Eq.~(\ref{eqk1}) into the condition~(\ref{cond}),
we obtain the ($n-1$)-dimensional FLRW-like equation as
\bea
\label{fr1}
\left( { \dot{r} \over r} \right)^2 \equiv H^2=-{1\over l^2}
+\left( {4\pi G_n \over n-2} \right)^2 \sigma^2-{1\over r^2}
+{\mu \over r^{n-1}}\,.
\eea
We choose $\sigma$ to cancel the cosmological constant
since we would like to consider a critical brane.

The standard FLRW equation without cosmological constant for 
an ($n-1$)-dimensional radiation-dominated universe is given by 
\bea
\label{frw1}
H^2 = -{1 \over r^2}+{16 \pi G_{n-1} \rho_b \over (n-2)(n-3)}\; .
\eea 
Here $\rho_b$ is the radiation energy density.
{}From the AdS/CFT point of view, the last term in Eq.~(\ref{fr1})
is interpreted as the contribution of the energy density of a CFT.
That is, we regard the energy density of an ($n-1$)-dimensional CFT
by means of the $n$-dimensional black hole mass parameter $\mu$ as
\bea
\rho_b = {(n-2)(n-3) \mu \over 16\pi G_{n-1} r^{n-1}}\,.
\label{rhob}
\eea 
Since the volume is given by $V_{n-1} = \Omega_{n-2} r^{n-2}$,  
the total energy on the brane takes the following form:
\bea
\label{en1}
E_{n-1}={(n-2)(n-3) \mu \Omega_{n-2}  \over 16\pi G_{n-1}\, r}  
=2M\,{l \over r}\; .
\eea 
This recovers the result given in Ref.~\cite{SV}, except for the
factor two. This difference is due to the fact that here we have
two copies of black holes as the bulk, i.e., $E_{bulk}=2M$. 

The Hawking temperature is 
\bea
\label{ht}
T_H={1\over \beta}=\left.{1\over 4\pi}
{d f(r)\over dr} \right|_{r=r_{+}} =
 {(n-1) r_+^4 +(n-3)r_+^2 l^2 
\over 4\pi  l^2 r_+^3  }\; .
\eea   
The temperature on the brane $T_{n-1}$ is given by the inverse of the 
periodicity of the proper time on the brane,
\bea
\label{t1}
T_{n-1}=T_H f(r)^{-1/2}\,.
\eea
The mass parameter $\mu$ may be expressed in terms of the
horizon radius $r_+$. From $f(r_+)=0$, we have
\bea
\label{mu}
\mu = \frac{r_+^{n-1}}{l^2} \left( 1+ \frac{l^2}{r_+^2}\right) \; .
\eea
Let us take the limit $r\gg l$ which is required for the
 AdS/CFT correspondence to hold,
so that $f(r)^{-1/2}$ is approximated by $l/r$. 

The entropy of an ($n-1$)-dimensional radiation bath is given by
\bea
\label{set}
S^{\rm RAD}_{n-1}={E_{n-1}+P_{n-1}V_{n-1} \over T_{n-1}}=
{(n-1)\over (n-2)} {E_{n-1} \over T_{n-1}}\,,
\eea 
where we have used the relation, $P_{n-1}V_{n-1}=E_{n-1}/(n-2)$.
Substituting Eqs.~(\ref{en1}), (\ref{ht}) and $f(r)^{-1/2}=l/r$
into the above, we obtain
\bea
\label{sn2}
S^{\rm RAD}_{n-1}&=&{(n-1)\over (n-2)}
{(n-2)(n-3) \Omega_{n-2}  \over 16\pi G_{n-1}\, r}f(r)^{1/2}  
 r_+^{n-3} \left( 1+ {r_+^2 \over l^2} \right) 
\left( {4\pi  l^2 r_+^3 \over (n-1) r_+^4 +(n-3)r_+^2 l^2 }\right) \nn
&=& {(n-3) \Omega_{n-2} r_+^{n-2} \over 4 G_{n-1} l }
\left(1+\frac{2}{n-1}\frac{l^2}{r_+^2}
+O\left(\frac{l^4}{r_+^4}\right)\right).
\eea
Note that this expression is independent of $r$.
Then, in the high temperature limit $r_+\gg l$,
we find $2S_n=S^{\rm RAD}_{n-1}$ from Eqs.~(\ref{sn1}) and (\ref{sn2}),
where again the factor two is due to the existence of the
two identical black holes.
Thus the entropy of the $n$-dimensional bulk and the
radiation entropy of the ($n-1$)-dimensional dual field theory 
coincide with each other in the high temperature limit.

\subsection{Einstein Gauss-Bonnet Gravity}

Now we consider Einstein Gauss-Bonnet gravity in the bulk.
The total action of the Einstein Gauss-Bonnet theory
is given by the sum of bulk and brane actions as before.
The bulk action is
\bea
\label{gb}
I_{\mbox{\scriptsize bulk}} =  {1\over 16\pi G_{n} } \int d^{n} x\sqrt{-g} 
\left( R -2\Lambda + \alpha {\cal L}_{\rm GB} \right) \; ,
\eea
where ${\cal L}_{\rm GB}$ is the Gauss-Bonnet term in Eq.~(\ref{gbt}).
Though the bulk equations become very complicated,
a simple black hole solution was found~\cite{BD}:
\bea
\label{metg}
ds^2 &=& -f(r)dt^2 + {1\over f(r)} dr^2 + r^2 d \Omega^2_{n-2}\;;
\\
\label{gbm}
f(r) &=& 1+{r^2 \over 2 \tilde{\alpha}} - {r^2 \over 2\tilde{\alpha}}
\sqrt{1+{4 \mu \tilde{\alpha} \over r^{n-1} } 
- {4 \tilde{\alpha}  \over \l^2}}\;,
\\
\tilde{\alpha}&=& (n-3)(n-4)\alpha
\nonumber\,.
\eea
Here $\mu$ is the mass parameter which is related to the black hole mass
in the same way as in Eq.~(\ref{mass}). The fact that the black hole
mass $M$ is equal to the total energy of the Einstein Gauss-Bonnet bulk 
can be checked by several methods~\cite{DT,DKO}.
In the limit $\alpha\to 0$, the metric~(\ref{metg}) reduces
to the metric of Einstein gravity in Eq.~(\ref{ein}).

The Hawking temperature is given by 
\bea
\label{ht2}
T_H&=&{1\over \beta}= {(n-1)r_+^4 +(n-3)r_+^2 l^2 
+ (n-5)\tilde{\alpha} l^2 \over 4\pi r_+ l^2 (r_+^2+2\tilde{\alpha})  }\; .
\eea
As in the case of Einstein gravity, the high temperature limit is 
given by $r_+ \gg l$.
Again, the entropy of the Einstein Gauss-Bonnet black hole can be 
obtained by the Euclidean method~\cite{MS},
\bea
\label{en2}
S_n = {\Omega_{n-2} r_{+}^{n-2}  \over 4 G_{n} }
\left( 1+ {2(n-2)\tilde{\alpha} \over (n-4)r_+^2 }\right)\; .
\eea
Note that this does not obey the area law unlike Einstein 
gravity.\footnote{A detailed derivation of the entropy by the
Euclidean method is given in Ref.~\cite{NP}. We can also obtain
Eq.~(\ref{en2}) by using a simple method based on 
thermodynamics of black holes~\cite{CAI}.}
As mentioned before, the Gauss-Bonnet term is a topological
invariant that does not enter the dynamics in $n=4$ dimensions.
Therefore, Eq.~(\ref{en2}) is valid for $n \ge 5$ dimensions.
 
The relationship between $G_n$ and $G_{n-1}$ takes the 
complicated form~\cite{GP,DS,MiS}:
\bea
\label{kank}
G_{n-1}&=&G_n  {(n-3) \over 2( 2- \sqrt{1-4\tilde{\alpha}/ l^2 } )}
\sqrt{ {1-\sqrt{1-4\tilde{\alpha}/ l^2 } \over 2\tilde{\alpha} }} \nn
&=&\frac{G_n (n-3) }{ 2 l}\,\left(1-\frac{3\tilde\alpha}{2l^2}
+O(\tilde\alpha^2)\right)\; ,
\eea
which reduces to Eq.~(\ref{ree}) in the limit $\alpha\to 0$.
Substituting the above relation to Eq.~(\ref{en2}), we obtain
the $n$-dimensional entropy in terms of the
 ($n-1$)-dimensional Newton constant $G_{n-1}$ as
\bea
\label{sn31}
S_n= {\Omega_{n-2} r_{+}^{n-2} (n-3)  \over 8  l G_{n-1} }
\left( 1-{3 \tilde{\alpha} \over 2 l^2 }
+ {2(n-2)\tilde{\alpha} \over (n-4)r_+^2 }
\right) + {\cal O}(\tilde{\alpha}^2) \; .
\eea
Here and in what follows, we assume $\tilde\alpha$ to be small
and retain only terms up to linear order in $\tilde\alpha$.
In the high temperature limit, the above formula gives
\bea
\label{sn3}
S_n \simeq{\Omega_{n-2} r_{+}^{n-2} (n-3)  \over 8  l G_{n-1} }
\left( 1-{3 \tilde{\alpha} \over 2 l^2 } \right).
\eea

We now consider the ($n-1$)-dimensional entropy on the brane.
As in the case of Einstein gravity,
the dynamics of the brane in the Gauss-Bonnet bulk is determined
by the junction condition at the brane~\cite{My2}. 
In contrast to Eq.~(\ref{eqk}), the junction condition
takes a very complicated form~\cite{SCD,GW} as 
\bea
\label{eqk2}
2({\cal K}_{ab}-{\cal K} h_{ab} )+2\tilde{\alpha}
\left( Q_{ab}-{1\over 3}Q h_{ab} \right) = -8\pi G_n S_{ab}\,,
\eea  
where
\bea
Q_{ab}&=&2 {\cal K}{\cal K}_{ac}{\cal K}^{c}_{~b}+
{\cal K}_{cd}{\cal K}^{cd}{\cal K}_{ab}
-2{\cal K}_{ac}{\cal K}^{cd}{\cal K}_{db}-{\cal K}^2{\cal K}_{ab}\nn
&&+2 {\cal K} R_{ab}+R{\cal K}_{ab}
-2{\cal K}^{cd}R_{cadb}-4R_{ac}{\cal K}_{b}^{~c}\; .
\eea 
Combining Eq.~(\ref{cond}) and the angular component of Eq.~(\ref{eqk2}),
we obtain the FLRW-like 
equation,\footnote{This was first derived by
Charmousis and Dufaux~\cite{CD}. Detailed calculations are
given in Refs.~\cite{SCD,GP}. In this paper, 
we adopt the notation of Ref.~\cite{GP}.} 
\bea
H^2&=&-{1\over r^2}+{c_+ + c_- -2 \over 4\tilde{\alpha}}\;;
\nn
c_{\pm}&=&\left(\sqrt{ 
\left\{ 1+{4 \mu \tilde{\alpha} \over r^{n-1} }
 - {4 \tilde{\alpha}  \over \l^2}
\right\}^{3/2} +9\left({4\pi G_n \over n-2}\sigma \right)^2 \tilde{\alpha} }
 \pm 3\left({4\pi G_n \over n-2}\sigma \right) 
{\tilde{\alpha}} \right)^{2/3} .
\eea
To first order in $\tilde\alpha$, we have
\bea
\label{fr3}
H^2&=&-{1\over l^2}
+\left( {4\pi G_n \over n-2} \right)^2 \sigma^2-{1\over r^2}
+{\mu \over r^{n-1}} \\
&&+\left( \left( -{1\over l^4} -{8\over 3}
\left( {4\pi G_n \over n-2} \right)^4 \sigma^4 \right)
-{\mu^2 \over r^{2n-2}}
+\left(-{4\mu \over r^{n-1}} + {4 \over l^2} \right)
\left( {4\pi G_n \over n-2} \right)^2 \sigma^2
+ {2\mu \over l^2 r^{n-1}} \right) \tilde{\alpha} \; .
\nonumber
\eea
Note that there is a term proportional to $\mu^2$, while
such a term is absent in Eq.~(\ref{fr1}).
Apparently, this term does not correspond to a radiation-like
energy density.
{}From a CFT perspective, this term may be regarded as
due to trace anomaly at next order in $1/N$ expansion.

We choose $\sigma$ in such a way that
the cosmological constant term in $H^2$ vanishes. 
The mass parameter $\mu$ is expressed in terms of 
the horizon radius $r_{+}$ as
\bea
\label{mu2}
\mu(r_{+}) = \frac{r_{+}^{n-1}}{l^2} 
\left( 1 + \frac{l^2}{r_{+}^2} + 
\frac{l^2\tilde{\alpha}}{r_{+}^4} \right) \; .
\eea
The energy density of the ($n-1$)-dimensional
field theory can be read from Eq.~(\ref{fr3}) as
\bea
\label{enn}
\rho_b
\simeq {(n-2)(n-3) \mu \over 16\pi G_{n-1}\, r^{n-1}}
\left( 1-{2 \tilde{\alpha}\over l^2} \right)\;+{\cal O}(\mu^2),
\eea
where we have taken the limit $r\gg r_+$ so as to remove
the unusual term proportional to $\mu^2/r^{2n-2}$. 
This limit is known as the near boundary limit discussed in Ref.~\cite{GP}.
 It should be noted that it was unnecessary to take such a limit 
for Einstein gravity, which is related to the existence of 
holography on the horizon. We come back to this point later.
Then, the radiation energy on the brane is
\bea
\label{pp}
E_{n-1}={(n-2)(n-3) \mu \Omega_{n-2}  \over 16\pi G_{n-1}\, r}
\left( 1-{2 \tilde{\alpha}\over l^2} \right)=
2M\,{l\over r}\left( 1-{2 \tilde{\alpha}\over l^2} \right)  \; .
\eea
It may be noted that the energy has a correction factor 
$(1-2\tilde\alpha/l^2)$. As seen from Eq.~(\ref{EGBredshift})
below, a quarter of it, that is $(1-\tilde\alpha/2l^2)$, is due to the correction in the redshift
factor, and the remain factor $(1-3\tilde\alpha/2l^2)$ is
due to the correction factor of the gravitational constant,
Eq.~(\ref{kank}).

The temperature on the brane $T_{n-1}$ is given by 
\bea
\label{tt}
T_{n-1}= T_H f(r)^{-1/2},
\eea
where the Hawking temperature $T_H$ is given by Eq.~(\ref{ht2}),
and $f(r)$ is  Eq.~(\ref{gbm}). 
Taking the limit $r\gg l$, and adopting 
the high temperature approximation, $r_+\gg l$, we have
\begin{eqnarray}
&&\beta=\frac{1}{T_H}
 \simeq \frac{4\pi l^2}{(n-1)r_+}\left(1+\frac{2\tilde\alpha}{r_+^2}
-\frac{n-3}{n-1}\frac{l^2}{r_+^2}\right)\,,
\nonumber\\
&&f(r)^{-1/2}\simeq\frac{l}{r}
\left(1+{\tilde{\alpha} \over l^2}\right)^{-1/2}
\simeq \frac{l}{r} \left(1- {\tilde{\alpha} \over 2 l^2}\right).
\label{EGBredshift}
\end{eqnarray}
The entropy of the ($n-1$)-dimensional radiation bath
is now obtained from eqs.~(\ref{pp}) and (\ref{tt}) as
\bea
\label{sn4}
S^{\rm RAD}_{n-1}&=& {\Omega_{n-2} r_{+}^{n-2} (n-3)  \over 4  l G_{n-1} }
\left( 1-{3 \tilde{\alpha} \over 2 l^2 }+ 2 {\tilde{\alpha} \over r_+^2}
+\frac{2}{n-1}\frac{l^2}{r_+^2}\right) 
+ {\cal O}(\tilde{\alpha}^2) \nn
&\simeq & {\Omega_{n-2} r_{+}^{n-2} (n-3)  \over 4  l G_{n-1} }
\left( 1-{3 \tilde{\alpha} \over 2 l^2 } \right) \;,  
\eea 
where the second line assumes the high temperature limit 
$r_+ \gg l$ again.
Thus we also obtain $2S_n=S^{\rm RAD}_{n-1}$
 from Eqs.~(\ref{sn3}) and (\ref{sn4})
in the high temperature limit.

\section{Hubble entropy}

\subsection{Einstein gravity}

As mentioned in the introduction, the Hubble entropy in Eq.~(\ref{sn5}) 
connects the Cardy-Verlinde entropy formula for CFT, which is a
generalization of the Cardy formula for CFT in 2 dimensions~\cite{CA}
to arbitrary dimensions, to the FLRW equation.
When the brane crosses the horizon $r=r_+$,
the FLRW equation (\ref{fr1}), together with Eq.~(\ref{mu}), gives
\bea
H^2={1\over l^2}\,.
\eea
Substituting $r=r_+$ and $H=1/l$ into Eq.~(\ref{sn5}),
we obtain
\bea
\label{sn7}
S^{\rm H}_{n-1}(r_+)= {(n-3) H V(r_+) \over 4 G_{n-1}}
={(n-3) r_+^{n-2} \Omega_{n-2}  \over 4 l G_{n-1}}\,.
\eea
Thus, comparing this with Eq.~(\ref{sn1}), we have 
$2S_n=S^{\rm H}_{n-1}(r_+)$.  Note that the agreement is
exact in the sense that there is no need to take the
high temperature limit. 

Furthermore, Verlinde demonstrated a very non-trivial
correspondence between the FLRW equation and a CFT~\cite{EV,SV}.
For the ($n-1$)-dimensional CFT side, the entropy 
is expressed in terms of the total energy and the Casimir energy 
as
\bea
\label{sn0}
S_{n-1}={2\pi r\over n-2}\sqrt{E_c (2E-E_c) }\; ,
\eea
where
\bea
\label{cas}
E_c \equiv (n-2)(E+PV-T_{n-1}S_{n-1})=
(n-2)T_{n-1}(S^{\rm RAD}_{n-1}-S_{n-1})\; .
\eea
Substituting into the above $S^{\rm RAD}_{n-1}$ in Eq.~(\ref{sn2}),
 $S_{n-1}=S^{\rm H}_{n-1}$, and $T_{n-1}$ in Eq.~(\ref{t1}),
the Casimir energy is identified as
\bea
E_c={(n-2)(n-3) \Omega_{n-2} r_+^{n-2} \over 8\pi G_{n-1} r_+ r} \;,
\eea
where the limit $r\gg l$ has been taken for $T_{n-1}$ in Eq.~(\ref{t1}).
It should be noted that the Casimir energy is negligible in the
high temperature limit as compared to the total energy $E$.
Verlinde showed that Eq.~(\ref{sn0}) is identical to the FLRW
equation evaluated at the horizon crossing, which can then be
generalized to arbitrary radius $r$.

\subsection{Einstein Gauss-Bonnet gravity}

Next, we consider the case of Einstein Gauss-Bonnet gravity. 
For the choice of the tension $\sigma$ to cancel the cosmological
constant, i.e., for a critical brane,
the FLRW equation~(\ref{fr3}) takes the form,
\begin{eqnarray}
H^2=-{1\over r^2}
+{\mu \over r^{n-1}} -\left(\frac{2\mu}{l^2r^{n-1}}
+\frac{\mu^2}{r^{2n-2}}\right) \tilde{\alpha} \; .
\label{EGBfr}
\end{eqnarray}
At the horizon crossing, $r=r_+$, this gives
\bea
\label{h23}
H^2={1\over l^2}
\left(1-\frac{3\tilde\alpha}{l^2}
-\frac{4\tilde\alpha}{r_+^2}\right)\,,
\eea
where Eq.~(\ref{mu2}) has been used.
Thus,
\bea
H \simeq {1\over l}
\left(1-\frac{3\tilde\alpha}{2l^2}
-\frac{2\tilde\alpha}{r_+^2}\right)\;,
\eea
and the Hubble entropy is given by
\bea
S^{\rm H}_{n-1}(r_+)={(n-3) H V(r_+) \over 4 G_{n-1}}
={(n-3) r_+^{n-2} \Omega_{n-2}  \over 4 l G_{n-1}} 
\left(1-\frac{3\tilde{\alpha}}{2l^2}-\frac{2\tilde\alpha}{r_+^2}\right)\;.
\eea
Taking the high temperature limit $r_+ \gg l$, 
we see that this coincides with $2S_n$ in Eq.~(\ref{sn3}),
\bea
\label{s9}
S^{\rm H}_{n-1}\simeq {(n-3) r_+^{n-2} \Omega_{n-2}  \over 4 l G_{n-1}} 
\left(1-{3\tilde\alpha\over 2 l^2}\right) \;.
\eea

One could have naively anticipated the above result
because the violation of the area law of the black hole entropy
is of $O(\tilde\alpha/r_+^2)$, which suggests the violation of
holography to be only of the same order of magnitude.
However, if we trace back the origin of the factor
$(1-3\tilde\alpha/2l^2)$ in Eq.~(\ref{s9}), we find that 
a part of it comes from the term $\mu^2/r_+^{2n-2}$ in the
FLRW equation (\ref{EGBfr}) that violates the conformal
invariance. This result suggests that there may be a way
to recover the exact holography in Einstein Gauss-Bonnet
gravity that supposedly takes account of next-to-leading order
corrections in $1/N$ expansion (i.e., 1-loop quantum corrections),
although we should not expect the Cardy-Verlinde formula
to hold, since it assumes the exact conformal invariance.

In this respect, we can appeal to a rather ad-hoc but
possibly interesting argument to recover holography in the
Einstein Gauss-Bonnet bulk. It is as follows.
Instead of adjusting the tension $\sigma$ to cancel
the cosmological constant, we consider a non-critical brane
and see if one can make the Hubble entropy to coincide
exactly with $2S_n$ up to corrections of $O(\tilde\alpha/r_+^2)$.
Interestingly, this turns out to be possible by allowing
a small positive cosmological constant on the brane.
Namely, we choose the brane tension $\sigma$ as
\bea
\left( {4\pi G_n \over n-2} \right)^2 \sigma^2 ={1\over l^2}
+\left(-{1\over 3l^4}
+{8(n-3) \over l^2 r_+^2 (n-4)} \right) \tilde{\alpha} \; .
\eea
Then, the FLRW equation becomes
\bea
\label{fr9}
H^2=-{1\over r^2}
+{\mu \over r^{n-1}}-\left({2\mu \over l^2 r^{n-1}}
+{\mu^2 \over r^{2n-2}}\right)\tilde\alpha
+ {8 (n-3) \over l^2 r_+^2 (n-4) }\,\tilde{\alpha} \;, 
\eea
where the last term corresponds to the cosmological constant on the brane.
At the horizon crossing, this gives
\bea
\label{h24}
H^2&=&-{1\over r_+^2}
+{\mu \over r_+^{n-1}}+\left( -{\mu^2 \over r_+^{2n-2}}-
{2\mu \over l^2 r_+^{n-1}} + {8 (n-3) \over l^2 r_+^2 (n-4) }\right) \tilde{\alpha} \nn
&=&{1\over l^2}-{1\over l^4}\left( 3-{4 (n-2) l^2\over (n-4) 
r_+^2} \right)\tilde{\alpha}+{\cal O}(\tilde{\alpha}^2)\;,
\eea
and the Hubble entropy is now given by
\bea
\label{snn}
S^{\rm H}_{n-1}(r_+)={(n-3) H V(r_+) \over 4 G_{n-1}}=
{(n-3) r_+^{n-2} \Omega_{n-2}  \over 4 l G_{n-1}} 
\left(1-{3\tilde{\alpha} \over 2l^2}
+ {2 (n-2) \tilde{\alpha} \over (n-4) r_+^2} \right) \;.
\eea
This agrees exactly with Eq.~(\ref{sn31}), namely, $2S_n=S^{\rm H}_{n-1}$
to $O(\tilde\alpha/r_+^2)$.

\section{Discussion}

In this paper, we discussed relations between 
the entropy of a black hole in $n$-dimensions and the entropy of 
its dual ($n-1$)-dimensional field theory
 for both Einstein and Einstein Gauss-Bonnet gravities.
By using the method in Ref.~\cite{GS}, we found the equivalence of 
the entropies in the high temperature limit $r_+ \gg l$
not only for Einstein but also for Einstein Gauss-Bonnet
to first order in the Gauss-Bonnet coupling parameter 
$\tilde{\alpha}$.

Contrary to the claim made in \cite{GP} that
there exists no holography for Einstein Gauss-Bonnet gravity
on the horizon,  we found that holography holds at least in the
high temperature limit, namely, the Hubble entropy at the
horizon crossing, $r =r_+$, coincides with the
black hole entropy in the limit $r_+\gg l$, where $r_+$ is
the horizon radius and $l$ is the AdS curvature radius.
 Interestingly, this coincidence was found
to be partly due to the existence of a term in the FLRW equation 
that violates the conformal invariance. We then argued,
though in an admittedly ad-hoc way, that holography may persists even to 
corrections of $O(\tilde\alpha/r_+^2)$, with an introduction of a small 
positive cosmological constant on the brane.

{}From a dual CFT perspective, Einstein Gauss-Bonnet gravity 
is considered to be the one that takes account of next-to-leading 
order corrections in $1/N$ expansion.
The fact that the entropy of an Einstein Gauss-Bonnet black hole 
and the CFT entropy induced on the brane are equal
in the high temperature limit implies that some part of
$1/N$ corrections simply contributes to redefinition/renormalization
of the tree level quantities, which seems 
quite a natural phenomenon for any field theory.

Of course, physically important information is contained in the 
remaining part of $1/N$ corrections, for which we are unable to
obtain any clue at the moment. As we mentioned in the above, 
one possibility is to assume the persistence of holography to
next-to-leading order in $1/N$ expansion. 
Then by deriving its consistency conditions,
we may be able to find some useful information.
For example, the small cosmological constant induced on the brane
may be interpreted as a result of the next-order (1-loop) quantum
corrections.  We hope to come back to this issue in the near future.

\section*{Acknowledgement}
S.O. thanks Prof. Nathalie Deruelle for helpful discussions on
the calculation of the entropy in Gauss-Bonnet gravity 
by the Euclidean method. 
The research of S.O. is supported by the Japan Society for the Promotion
of Science. This work is supported in part by Monbukagaku-sho
Grant-in-Aid for Scientific Research (S) No.\ 14102004.


\begin{thebibliography}{99}

\bibitem{RS2} L.~Randall, R.~Sundrum, 
``An Alternative to Compactification,''
Phys.\ Rev.\ Lett. {\bf 83}, 4690 (1999) 3370 
[arXiv:hep-th/9906064]. 

\bibitem{Gubser:1999vj}
S.~S.~Gubser,
``AdS/CFT and gravity,''
Phys.\ Rev.\ D {\bf 63}, 084017 (2001)
[arXiv:hep-th/9912001].

\bibitem{GS}
J.~Garriga and M.~Sasaki,
``Brane-world creation and black holes,''
Phys.\ Rev.\ D {\bf 62}, 043523 (2000)
[arXiv:hep-th/9912118].

\bibitem{SV} I.~Savonije and E.~Verlinde,
``CFT and entropy on the brane,''
Phys.\ Lett.\ B {\bf 507},305 (2001) [arXiv:hep-th/0102042].
 
\bibitem{GP1} A.~Padilla,
``CFTs on noncritical Brane Worlds,''
Phys.\ Lett.\ B {\bf 528}, 274 (2002)
[arXiv:hep-th/0111247].

\bibitem{GP2} J.~P.~Gregory and A.~Padilla,
``Exact brane world cosmology induced from bulk black holes,''
Class.\ Quant.\ Grav.\ {\bf 19} 4071, (2002)
[arXiv:hep-th/0204218].

\bibitem{GP3} A.~Padilla,
``Brane world cosmology and holography,''
Ph.D. Thesis (Advisor: Ruth Gregory).
[arXiv:hep-th/0210217]. 

\bibitem{Maldacena:1997re}
J.~M.~Maldacena,
``The large N limit of superconformal field theories and supergravity,''
Adv.\ Theor.\ Math.\ Phys.\  {\bf 2}, 231 (1998)
[Int.\ J.\ Theor.\ Phys.\  {\bf 38}, 1113 (1999)]
[arXiv:hep-th/9711200].

\bibitem{EW} E.~Witten, 
``Anti-de Sitter Space, Thermal Phase Transition, 
And Confinement In Gauge Theories,'' 
Adv.\ Theor.\ Math.\ Phys.\ {\bf 2}, 505 (1998) 
[arXiv:hep-th/9803131].

\bibitem{EV} E.~Verlinde, 
``On the holographic principle in a radiation dominated universe,''
[arXiv:hep-th/0008140].

\bibitem{CA} J.~L.~Cardy, 
``Operator content of two-dimensional conformally invariant theories,''  
Nucl.\ Phys.\ B {\bf 270},186 (1986)


\bibitem{DL} D.~Lovelock, 
``The Einstein tensor and its generalizations,''
J.\ Math.\ Phys.\ {\bf 12}, 498(1971).


\bibitem{CH} P.~Candelas and G.~T.~Horowitz, A.~Strominger and E.~Witten, 
``Vacuum configurations for superstrings,'' 
Nucl.\ Phys.\ B {\bf 258}, 46 (1985).


\bibitem{BZ} B.~Zwiebach, 
``Curvature squared terms and string theories,''
Phys.\ Lett.\ B {\bf 156}, 315(1985).


\bibitem{BZU} B.~Zumino, 
``Gravity theories in more than four-dimensions,''
Phys.\ Rept.\ {\bf 137},109 (1986).

\bibitem{GSl} D.~J.~Gross and J.~H.~Sloan,
``The quartic effective action for the heterotic string,'' 
Nucl.\ Phys.\ B {\bf 291},41 (1987).

\bibitem{NO} S.~Nojiri and S.~D.~Odintsov, 
``Brane-world cosmology in higher derivative gravity or warped compactification in the next-to-leading order of AdS/CFT correspondence,''
JHEP\ {\bf 07}, 049 (2000) [arXiv:hep-th/0006232].

\bibitem{MS} R.~C.~Myers, J.~Z.~Simon, 
``Black-hole thermodynamics in Lovelock gravity,''
Phys.\ Rev.\ D  {\bf 38}, 2434(1988).

\bibitem{SN1} S.~Nojiri, S.~D.~Odintsov,
``Anti-de Sitter Black Hole Thermodynamics in Higher Derivative Gravity 
and New Confining-Deconfining Phases in dual CFT,''
Phys.\ Lett.\ B {\bf 521}, 87 (2001) [arXiv:hep-th/0109122].

\bibitem{CAI} R.-G.~Cai, 
``Gauss-Bonnet black holes in AdS spaces,''
Phys.\ Rev.\ D {\bf 65}, 084014 (2002) [arXiv:hep-th/0109133].

\bibitem{SN2} M.~Cvetic, S.~Nojiri and S.~D.~Odintsov,
``Black Hole Thermodynamics and Negative Entropy in deSitter 
and Anti-deSitter Einstein-Gauss-Bonnet gravity,''
Nucl.\ Phys.\ B {\bf 628}, 295 (2002) [arXiv:hep-th/0112045].

\bibitem{NP2} I.~P.~Neupane, 
``Black hole entropy in string-generated gravity models,''  
Phys.\ Rev.\ D {\bf 67}, 061501(2003)
[arXiv:hep-th/0212092 ]. 

\bibitem{CAI2} R.-G.~Cai,  Q.~Guo,
``Gauss-Bonnet Black Holes in dS Spaces,''
Phys.\ Rev.\ D {\bf 69}, 104025 (2004) [arXiv:hep-th/0311020 ].

\bibitem{CAI3} R.-G.~Cai, 
``A Note on Thermodynamics of Black Holes in Lovelock Gravity,''
Phys.\ Lett.\ B {\bf 582}, 237  (2004) [arXiv:hep-th/0311240].

\bibitem{SN3} S.~Nojiri, S.~D.~Odintsov and S.~Ogushi,
``Holographic entropy and brane FRW-dynamics from AdS black hole 
in d5 higher derivative gravity,''
Int.\ J.\ Mod.\ Phys.\ A {\bf 16}, 5085 (2001) [arXiv:hep-th/0105117 ].

\bibitem{SN4} S.~Nojiri, S.~D.~Odintsov and S.~Ogushi,
``Cosmological and black hole brane-world Universes in higher derivative 
gravity,'' 
Phys.\ Rev.\ D {\bf 65}, 023521 (2002) [arXiv:hep-th/0108172]. 

\bibitem{NP} Y.~M.~Cho and I.~P.~Neupane, 
``Anti-de Sitter black holes, thermal phase transition, and holography 
in higher curvature gravity,''
Phys.\ Rev.\ D {\bf 66},024044 (2002) [arXiv:hep-th/0202140].

\bibitem{SN5} S.~Nojiri, S.~D.~Odintsov and S.~Ogushi,
``Friedmann-Robertson-Walker brane cosmological equations from the 
five-dimensional bulk (A)dS black hole,''
Int.\ J.\ Mod.\ Phys.\ A {\bf 17}, 4809 (2002) [arXiv:hep-th/0205187].

\bibitem{CAI1} R.-G.~Cai, Y.~S.~ Myung
``Holography and Entropy Bounds in Gauss-Bonnet Gravity,''
Phys.\ Lett.\ B {\bf 559}, 60 (2003) [arXiv:hep-th/0210300].

\bibitem{GP} J.~P.~Gregory and A.~Padilla, 
``Braneworld holography in Gauss-Bonnet gravity,''
Class.\ Quant.\ Grav.\ {\bf 20} 4221, (2003)
[arXiv:hep-th/0304250].

\bibitem{GibHaw77}  G.~W.~Gibbons and S.~W.~Hawking, 
``Action integrals and partition functions in quantum gravity,''
Phys.\ Rev.\ D {\bf 15}, 2752 (1977) 

\bibitem{HawPag83} S.~W.~Hawking and D.~N.~Page, 
``Thermodynamics of black holes in Anti-de Sitter space,''
Commun.\  Math.\ Phys.\ {\bf 87}, 577 (1983).

\bibitem{ADM} R.~Arnowitt,~S.~Deser and C.~Misner, 
``Dynamical Structure and Definition of Energy in General Relativity,''
Phys.\ Rev.\ {\bf 116}, 1322 (1959); 
``Canonical Variables for General Relativity,''
Phys.\ Rev.\ {\bf 117}, 1595 (1960) 1595.


\bibitem{AD} L.~F.~Abbott and S.~Deser, 
``Stability of gravity with a cosmological constant,''  
Nucl.\ Phys.\ B {\bf 195}, 76 (1982).

\bibitem{KBL} J.~Katz, J.~Bicak, D.~Lynden-Bell, 
``Relativistic conservation laws and integral constraints for 
large cosmological perturbations,''
Phys.\ Rev.\ D {\bf 55}, 5957 (1997).

\bibitem{GKR} S.~B.~Giddings, E.~Katz, and L.~Randall,
``Linearized gravity in brane backgrounds,''
JHEP\ {\bf 03},023 (2000) [arXiv:hep-th/0002091].

\bibitem{BD} D.~G.~Boulware and S.~Deser,
``String-Generated Gravity Models,''
Phys.\ Rev.\ Lett.\ {\bf 55}, 2656 (1985).

\bibitem{DT} S.~Deser and B.~Tekin, 
``Energy in generic higher curvature gravity theories,''
Phys.\ Rev.\ D {\bf 67}, 084009 (2003)  
[arXiv:hep-th/0212292].

\bibitem{DKO} N.~Deruelle, J.~Katz and S.~Ogushi,
``Conserved charges in Einstein Gauss-Bonnet theory,''
Class.\ Quant.\ Grav.\ {\bf 21},1971 (2004)
[arXiv:gr-qc/0310098].

\bibitem{DS} N.~Deruelle and M.~Sasaki, 
``Newton's Law on an Einstein gGauss-Bonnet" Brane,''
Prog.\ Theor.\ Phys.\ {\bf 110}, 441 (2003) [arXiv:gr-qc/0306032]. 

\bibitem{MiS} M.~Minamitsuji and M.~Sasaki,
``Linearized gravity on the de Sitter brane 
in the Einstein Gauss-Bonnet theory,'' [arXiv:hep-th/0404166].

\bibitem{My2} R.~C.~Myers, 
``Higher-derivative gravity, surface terms, and string theory,''
Phys.\ Rev.\ D {\bf 36},392 (1987).

\bibitem{SCD} S.~C.~Davis, 
``Generalized Israel junction conditions for a Gauss-Bonnet brane world,''
Phys.\ Rev.\ D {\bf 67},  024030 (2003) 
[arXiv:hep-th/0208250]. 

\bibitem{GW} E.~Gravanis and S.~Willison, 
``Israel conditions for the Gauss-Bonnet theory and the 
Friedmann equation on the brane universe,''  
Phys.\ Lett.\ B {\bf 562},118 (2003) 
[arXiv:hep-th/0209076].

\bibitem{CD} C.~Charmousis and J.~-F.~ Dufaux, 
``General Gauss-Bonnet brane cosmology,''
Class.\ Quant.\ Grav.\ {\bf 19},4671 (2002)
[arXiv:hep-th/0202107]. 




\end{thebibliography}
\end{document}